\documentclass[fleqn,10pt]{wlscirep}
\usepackage{amssymb,amscd,amsmath}
\usepackage{graphicx}
\usepackage{cite}
\usepackage{array}
\newcolumntype{L}[1]{>{\raggedright\let\newline\\\arraybackslash\hspace{0pt}}m{#1}}
\newcolumntype{C}[1]{>{\centering\let\newline\\\arraybackslash\hspace{0pt}}m{#1}}
\newcolumntype{R}[1]{>{\raggedleft\let\newline\\\arraybackslash\hspace{0pt}}m{#1}}
\def\hsp{\hspace}

\def\tm{truemm}
\def\a={&=&}
\def\aa={&\approx&}
\def\ad={&\equiv&}
\def\ap={&\propto&}
\def\as= {&\sim&}
\def\lra={&\Leftrightarrow&}
\def\sa={\hsg &=& \hsg}
\def\saa={\hsg &\approx&\hsg}
\def\sad={\hsg &\equiv&\hsg}
\def\sap={\hsg &\propto&\hsg}
\def\sass={\hsg &\sim&\hsg}
\def\slra={\hsg &\Leftrightarrow&\hsg}
\def\b={\ = \ }
\def\ba={\ \approx \ }
\def\bp={\ \propto \ }
\def\c={\hsa = \hsa}
\def\ca={\hsa \approx \hsa}
\def\abs= {\ \sim \ }
\def\acs= {\hsa \sim \hsa}

\def\bHu{\begin{Huge}}
\def\bhu{\begin{huge}}
\def\bLA{\begin{LARGE}}
\def\bLa{\begin{Large}}
\def\bla{\begin{large}}
\def\bsm{\begin{small}}
\def\bft{\begin{footnotesize}}
\def\bsc{\begin{scriptsize}}
\def\bti{\begin{tiny}}
\def\eHu{\end{Huge}}
\def\ehu{\end{huge}}
\def\eLA{\end{LARGE}}
\def\eLa{\end{Large}}
\def\ela{\end{large}}
\def\esm{\end{small}}
\def\eft{\end{footnotesize}}
\def\esc{\end{scriptsize}}
\def\eti{\end{tiny}}
\def\barr{\begin{array}}
\def\earr{\end{array}}
\def\bc{\begin{center}}
\def\ec{\end{center}}
\def\ben{\begin{enumerate}}
\def\een{\end{enumerate}}
\def\beq{\begin{equation}}
\def\eeq{\end{equation}}
\def\beqa{\begin{eqnarray}}
\def\eeqa{\end{eqnarray}}
\def\beqanu{\vspace{-7mm} \begin{eqnarray*}}
\def\beqan{\begin{eqnarray*}}
\def\eeqan{\end{eqnarray*}}
\def\bit{\begin{itemize}}
\def\eit{\end{itemize}}
\def\bpm{\begin{pmatrix}}
\def\epm{\end{pmatrix}}
\def\bvs{\begin{verse}}
\def\evs{\end{verse}}
\def\btab{\begin{tabular}}
\def\etab{\end{tabular}}

\def\hsa{\mbox{} \hspace{2\tm} }
\def\hsb{\mbox{} \hspace{4\tm} }

\def\hsg{\mbox{} \hspace{-2\tm} }

\newcommand{\pa}[1]{\left( {#1} \right)}

\newcommand{\pas}[1]{\left[ {#1} \right]}

\newcommand{\paf}[2]{\left( \frac{#1}{#2} \right)}

\title{Comparing reactive and memory-one strategies of direct reciprocity}

\author[1]{Seung Ki Baek}
\author[2,$\ast$]{Hyeong-Chai Jeong}
\author[3,4,$\dagger$]{Christian Hilbe}
\author[4,$\ddagger$]{Martin A. Nowak}
\affil[1]{Department of Physics, Pukyong National University, Busan 48513,
Korea}
\affil[2]{Department of Physics and Astronomy, Sejong University, Seoul
05006, Korea}
\affil[3]{IST Austria, Am Campus 1, 3400 Klosterneuburg, Austria}
\affil[4]{Program for Evolutionary Dynamics,
Department of Mathematics, Harvard University, Cambridge, MA 02138, United
States of America}

\affil[$\ast$]{hcj@sejong.edu}
\affil[$\dagger$]{christian.hilbe@ist.ac.at}
\affil[$\ddagger$]{martin\_nowak@harvard.edu}

\begin{abstract}
\noindent
Direct reciprocity is a mechanism for the evolution of cooperation based on
repeated interactions. When individuals meet repeatedly, they can use
conditional strategies to enforce cooperative outcomes that would not be
feasible in one-shot social dilemmas. Direct reciprocity requires that
individuals keep track of their past interactions and find the right response.
However, there are natural bounds on strategic complexity: Humans find it
difficult to remember past interactions accurately, especially over long
timespans. Given these limitations, it is natural to ask how complex strategies
need to be for cooperation to evolve. Here, we study stochastic evolutionary
game dynamics in finite populations to systematically compare the evolutionary
performance of reactive strategies, which only respond to the co-player’s
previous move, and memory-one strategies, which take into account the own and
the co-player’s previous move. In both cases, we compare deterministic strategy
and stochastic strategy spaces. For reactive strategies and small costs, we find
that stochasticity benefits cooperation, because it allows for
generous-tit-for-tat. For memory one strategies and small costs, we find that
stochasticity does not increase the propensity for cooperation, because the
deterministic rule of win-stay, lose-shift works best. For memory one strategies
and large costs, however, stochasticity can augment cooperation.
\end{abstract}
\begin{document}

\flushbottom
\maketitle
\thispagestyle{empty}

\section*{Introduction}
Direct reciprocity, the propensity to return cooperative acts of others, is one
of the major mechanisms to establish cooperation
\cite{trivers:qrb:1971,axelrod:book:1994,nowak:science:2006}. The theory of
reciprocity has allowed us to understand under which conditions ``a shadow of
the future'' can help individuals to forego individual short-run benefits in
favour of mutually beneficial long-run relationships \cite{nowak:book:2006,
sigmund:book:2010,fudenberg:eco:1986, boyd:jtb:1988, hauert:prsb:1997,
pacheco:jtB:2008, rand:jtb:2009, vanveelen:pnas:2012,bednarik:prsb:2014,
szolnoki:scirep:2014}.
Although reciprocal relationships also seem to be at work in several animal
species \cite{wilkinson:nature:1984,milinski:nature:1987,
stephens:science:2002}, they play a particular role for human interactions
\cite{binmore:book:2011}. Because almost all our social interactions occur
repeatedly, reciprocity considerations may have played an important role for the
evolution of social heuristics \cite{rand:natcomms:2014, capraro:scirep:2014},
which in turn helps to understand why we also cooperate with strangers
\cite{delton:pnas:2011}, sometimes even without considering the resulting costs
to ourselves \cite{hoffman:pnas:2015}.

To model the emergence of direct reciprocity, researchers often use the example
of the iterated prisoner's dilemma. In this game, two players can decide
repeatedly whether to cooperate or to defect. While mutual cooperation is
optimal from a group perspective, players may feel a temptation to defect at the
expense of the co-player. Strategies for the repeated prisoner's dilemma can
become arbitrarily complex -- sophisticated players may use the whole past
history of play when making the decision whether to cooperate in the next round.
In practice, however, several experiments suggest that the complexity of human
strategies is restricted. For example, Stevens et al. \cite{stevens:fip:2011}
have shown that subjects have problems to remember their co-players' past
decisions accurately, especially if they need to keep track of several
co-players or multiple rounds. Similarly, the research of Wedekind and Milinski
\cite{milinski:pnas:1998,wedekind:science:2000} suggests that there is a
trade-off between having a
sophisticated strategy in the prisoner's dilemma and performing well in a second
unrelated task. In addition, recent studies in behavioural economics have found
that most of the strategies employed by human subjects are well described by
simple strategies that only depend on the last interaction
\cite{engle:et:2006,dalbo:aer:2011,camera:geb:2012,bruttel:td:2012,dalbo:ssrn:2015},
although there are also other factors such as the average fraction of past
cooperative acts \cite{cuesta:scirep:2015,gallo:pnas:2015}.
Given that there are such constraints on the complexity of
strategies, can we still expect cooperation to evolve? And how complex do the
players' strategies need to be in order to allow for substantial cooperation?

Herein, we approach this question by comparing the evolving cooperation rates
for different strategy spaces for the repeated prisoner's dilemma. The
considered strategy spaces differ along two dimensions of complexity. The first
dimension is the input that they require:
whereas reactive strategies (or memory-1/2
strategies) only require information about the co-player's previous
move~\cite{kalai:ijgt:1988,nowak:jtb:1989,wahl:jtb:1999},
memory-one strategies additionally need to take one's own move into
account~\cite{nowak:pnas:1993}. The set of reactive strategies is a reasonable
and conventional choice to define a subset of memory-one strategies,
because a player's payoff crucially depends on the co-player's move in
the prisoner's dilemma.
The second dimension is the strategy's stochasticity. Here, we distinguish
strategies that respond to past outcomes in a deterministic fashion, and
strategies that prescribe to randomize. Overall, these two independent
dimensions of complexity lead to four different strategy classes.

To assess whether a given strategy class is favourable to the evolution of
cooperation, we consider the Moran process in a finite population of players
\cite{nowak:nature:2004}. Individuals can choose freely among the available
strategies, and over time they learn to switch to strategies that yield a higher
payoff. By assuming that mutations are sufficiently rare, we can use the
framework of Fudenberg \& Imhof \cite{fudenberg:jet:2006} to calculate how
often players use each of the available strategies in the long run
\cite{martinez:plosone:2012}. This in turn allows us to calculate the evolving
cooperation rates for each of the four strategy classes, as explained in more
detail in the next section. Our results suggest that strategies with larger
memory are typically beneficial for the evolution of cooperation, whereas the
strategies' stochasticity can sometimes have a detrimental effect.

\section*{Model and Methods}

It is common to consider two levels when modelling the evolutionary dynamics of
repeated games. The first level focuses on the repeated game itself. At this
level, we look at a single instance of the repeated game and we calculate how
the players' strategies determine the resulting cooperation rates and average
payoffs. The second level describes the population dynamics. Here, we look at a
whole population of players. Each player is equipped with a strategy for how to
play the repeated game. The abundance of a given strategy within the population
may change over time, because strategies that lead to a high payoff are expected
to spread (either due to reproduction of successful individuals, or due to
imitation and cultural learning). At the population level, we are interested in
how often a strategy will be used in the long run, and what the resulting
average cooperation rate is. In the following, we describe these two levels in
more detail.

\subsection*{Game dynamics of the repeated prisoner's dilemma}

In the prisoner's dilemma, two individuals decide simultaneously whether to
cooperate ($C$) or to defect ($D$). A player who cooperates pays a cost $c>0$ to
provide a benefit $b>c$ for the co-player. Thus, a cooperator either gets $b-c$
(if the co-player cooperates as well) or $-c$ (if the co-player defects). On the
other hand, a defector either gets $b$ (if the co-player cooperates) or $0$ (if
the co-player defects). To reduce the number of free parameters, we can set
$b:=1$ and we let $c$ vary between $0\!<\!c\!<\!1$. Moreover, to avoid negative
payoffs, we add the constant $c$ to all payoffs. Under these assumptions, the
payoff matrix of the prisoner's dilemma takes the form
\beqa
\hsp{-10\tm}
\begin{array}{cl}
 \hsb   & \begin{array}{cc}
     \hsa\hsb\hsa  C  &\hsb\hsa D
       \end{array} \ \	\\
\begin{array}{c}
      C      \\
      D
\end{array}
   & \left( \begin{array}{ccc}
\hsa    1   & \hsb   0 \hsb\hsa\\
\hsa    1+c     & \hsb   c\hsb\hsa
  \end{array} \right).
\end{array}
\label{e.CD}
\eeqa
Because $c\!<\!1$, both players prefer mutual cooperation over mutual defection;
however, since $c\!>\!0$, each individual is tempted to play $D$ irrespective of
the co-player's action. If the prisoner's dilemma is played in a well-mixed
population, evolution favours defection.

The question of evolutionary strategy selection becomes more interesting when
individuals have the option to reciprocate past actions in the future. To model
such repeated interactions, we consider two individuals who play the game
(\ref{e.CD}) for infinitely many rounds. Strategies for such repeated games need
to prescribe an action for any possible history of previous play, and they can
become arbitrarily complex. To facilitate an evolutionary analysis, we assume
herein that individuals at most make use of simple memory-one strategies. That
is, their behaviour in any given round may only depend on the outcome of the
previous round. Memory-one strategies can be written as a 4-tuple,
$\mathbf{p}=(p_{CC},p_{CD},p_{DC},p_{DD})$. The entries $p_{ij}$ correspond to
the player's probability to cooperate in the next round, given that the focal
player's previous action was $i$ and that the co-player's action was $j$. We
assume that players only have imperfect control over their actions, such that
they mis-implement their intended action with some small probability
$\varepsilon>0$ \cite{boyd:jtb:1989,sigmund:book:2010}. Under this assumption,
the player's effective strategy becomes
$\mathbf{p}'=(1-\varepsilon)\mathbf{p}+\varepsilon (\mathbf{1}-\mathbf{p})$.

When both players apply memory-one strategies $\mathbf{p}$ and $\mathbf{q}$,
respectively, then the dynamics of the repeated prisoner's dilemma takes the
form of a Markov chain with four possible states $CC$, $CD$, $DC$, $DD$ (the
possible outcomes of each round). The transition matrix of this Markov chain is
given by
\begin{equation}
\begin{pmatrix}
p'_{CC} q'_{CC} & p'_{CC}(1- q'_{CC}) & (1-p'_{CC}) q'_{CC} & (1-p'_{CC})(1-
q'_{CC}) \\
p'_{CD} q'_{DC} & p'_{CD} (1-q'_{DC}) & (1-p'_{CD}) q'_{DC} &
(1-p'_{CD})(1-q'_{DC}) \\
p'_{DC} q'_{CD} & p'_{DC} (1-q'_{CD}) & (1-p'_{DC}) q'_{CD} & (1-p'_{DC})
(1-q'_{CD}) \\
p'_{DD} q'_{DD} & p'_{DD} (1-q'_{DD}) & (1-p'_{DD}) q'_{DD} & (1-p'_{DD})
(1-q'_{DD})
\end{pmatrix}.
\end{equation}
Due to the assumption of errors, all entries of this transition matrix are
positive. Therefore, there exists a unique invariant distribution
$\mathbf{v}=(v_{CC},v_{CD},v_{DC},v_{DD})$, representing the probability to find
the two players in each of the four states over the course of the game. Given
the invariant distribution $\mathbf{v}$, we can calculate player 1's payoff as
$\pi(\mathbf{p},\mathbf{q})=\mathbf{v} \cdot \mathbf{h}_1$ and player 2's payoff
as $\pi(\mathbf{q},\mathbf{p})=\mathbf{v} \cdot \mathbf{h}_2$, with
$\mathbf{h}_1=(1,0,1+c,c)$ and $\mathbf{h}_2=(1,1+c,0,c)$. Similarly, we can
calculate the players' average cooperation rate in the repeated game as
$\gamma(\mathbf{p},\mathbf{q})=v_{CC}\!+\!v_{CD}$ and
$\gamma(\mathbf{q},\mathbf{p})=v_{CC}\!+\!v_{DC}$. If the cooperation rate
$\gamma(\mathbf{p},\mathbf{p})$ of a strategy against itself converges to one as
the error rate $\varepsilon $ goes to zero, we call the strategy $\mathbf{p}$ a
{\it self-cooperator} (see also ref.~\citeonline{stewart:pnas:2014}). Similarly,
strategies for which the cooperation rate $\gamma(\mathbf{p},\mathbf{p})$
approaches zero are called {\it self-defectors}.

We are interested in how the complexity of the strategy space affects the
evolution of cooperation. To this end, we distinguish two dimensions of
complexity. The first dimension is the input that the strategy takes into
consideration. Players with a
memory-1 strategy take the full outcome of the previous round into account,
whereas players with a reactive strategy (or memory-1/2 strategy) only consider
the co-player's previous move (but not the own move). The second dimension is
the strategy's stochasticity. Players with a deterministic strategy respond to
past outcomes in a deterministic fashion, whereas players with a stochastic
strategy may randomize between cooperation and defection. Combining these two
dimensions, we end up with four different strategy spaces, as summarized in
Table~\ref{table:class}.
\begin{table}
\bgroup
\def\arraystretch{2}
\begin{tabular}{ C{2.6cm}|C{6.7cm}|C{6.2cm}| }
\multicolumn{1}{r}{}
 & \multicolumn{1}{c}{{\bf Reactive strategies}}
 & \multicolumn{1}{c}{{\bf Memory-1 strategies}} \\[1ex]
\cline{2-3}
{\bf Deterministic strategies} & Deterministic reactive
strategies, $\mathcal{M}_{1/2}$
$p_{CC}\!=\!p_{DC}, ~p_{CD}\!=\!p_{DD}$
\hspace{5cm}$p_{ij}\in\{0,1\}$
& Deterministic memory-1 strategies, $\mathcal{M}_{1}$
$p_{ij}\in\{0,1\}$ \\
\cline{2-3}
{\bf Stochastic strategies} & Stochastic reactive
strategies, $\mathcal{\hat{M}}_{1/2}$
$p_{CC}\!=\!p_{DC}, ~p_{CD}\!=\!p_{DD}$
\hspace{5cm}$p_{ij}\in[0,1]$
& Stochastic memory-1 strategies, $\mathcal{\hat{M}}_{1}$ $p_{ij}\in[0,1]$ \\
\cline{2-3}
\end{tabular}
\egroup
\caption{Four different strategy spaces considered in this work. Each parameter
$p_{ij}$ denotes the focal player's probability to cooperate in the next round,
given that the player's previous action was $i$ and that the co-player's action
was $j$.}
\label{table:class}
\end{table}
~\\
These four strategy spaces are partially ordered, $\mathcal{M}_{1/2}\! \subseteq
\!\mathcal{M}_1 \! \subseteq \! \hat{\mathcal{M}}_1$ and $\mathcal{M}_{1/2} \!
\subseteq \! \hat{\mathcal{M}}_{1/2} \! \subseteq \! \hat{\mathcal{M}}_1$ (there
is no order between $\hat{\mathcal{M}}_{1/2}$ and $\mathcal{M}_1$). Examples of
deterministic reactive strategies include $AllD=(0,0,0,0)$, $AllC=(1,1,1,1)$ and
Tit-for-Tat, $TFT=(1,0,1,0)$. An example of a stochastic reactive strategy is
generous Tit-for-Tat, $GTFT=(1,1\!-\!c/b,1,1\!-\!c/b)$ (see
refs.~\citeonline{molander:jcr:1985,nowak:nature:1992}).
Finally, as two examples of deterministic memory-one
strategies which are not reactive, we mention the Grim Trigger strategy,
$GT=(1,0,0,0)$, and Win-stay Lose-shift, $WSLS=(1,0,0,1)$. $GT$ switches to
relentless defection after any deviation from mutual cooperation; $WSLS$, on the
other hand, sticks to an action if and only if it has been successful in the
previous round \cite{kraines:td:1989,nowak:nature:1993,tamura:epjd:2015}.

\subsection*{Population dynamics}

To describe the evolutionary dynamics on the population level, we use the Moran
process \cite{nowak:nature:2004,taylor:bmb:2004,nowak:book:2006,jeong:jtb:2014}
in the limit of rare mutations
\cite{fudenberg:jet:2006,wu:jmb:2012,mcavoy:jet:2015}. That is, we consider a
population of size $N$, and we suppose that new mutant strategies are
sufficiently rare such that at any moment in time at most two different
strategies are present in the population. If there are $i$ individuals who adopt
the strategy $\mathbf{p}$, and $N-i$ individuals who adopt the strategy
$\mathbf{q}$, the average payoffs for the two groups of players are
 \beqa
F_i 	\a= \frac{(i-1)\cdot \pi(\mathbf{p},\mathbf{p})+ (N-i)\cdot
\pi(\mathbf{p},\mathbf{q})}{N-1} \\[0.3cm]
G_i \a= \frac{i\cdot \pi(\mathbf{q},\mathbf{p}) +\pa{N-i-1}\cdot
\pi(\mathbf{q},\mathbf{q})}{N-1}.
 \eeqa
We assume that the fitness of a strategy is a linear function of its payoff.
Specifically, if the fitness of the strategies $\mathbf{p}$ and $\mathbf{q}$ is
denoted by $f_i$ and $g_i$, respectively, then
 \beqa
 f_i \a= 1 + w \cdot F_i \\[0.2cm]
 g_i \a= 1 + w \cdot G_i.
 \label{eq.fit}
 \eeqa
The constant terms on the right-hand side correspond to the player's background
fitness, and the parameter $w$ is a measure for the strength of selection. When
$w \rightarrow 0$, payoffs become irrelevant, and both strategies have
approximately equal fitness. We refer to this special case as the limit of weak
selection.

The abundance of a strategy can change over time, depending on the strategy's
relative success. We consider a simple birth-death process. In each time step,
one individual is randomly chosen for death, and its place is filled with the
offspring of another individual, which is randomly chosen proportional to its
fitness. That is, if $T_i^\pm$ denotes the probability that the number of
individuals with strategy $\mathbf{p}$ becomes $i\pm 1$ after one time step,
then we can calculate
 \beqa
  T_i^+ \a= \paf{i f_i }{i f_i + \pa{N-i}g_i}\paf{N-i}{N}\\[0.3cm]
  T_i^- \a= \paf{\pa{N-i} g_i }{i f_i + \pa{N-i}g_i}\paf{i}{N}.
  \label{e.Tnapm}
 \eeqa
The quantities $T_i^+$ and $T_i^-$ can be used to compute the probability that
eventually the whole population will adopt strategy $\mathbf{p}$
\cite{nowak:nature:2004}. In the special case that the population starts from a
state in which only a single player applies $\mathbf{p}$, this fixation
probability $\rho$ is given by
\beqa
 \rho(\mathbf{p},\mathbf{q}) \a= \pas{1+\sum_{j=1}^{N-1} \prod_{i=1}^j
\frac{T_i^-}{T_i^+}}^{-1}.
\label{e.fix}
\eeqa
If there is no selection (i.e., if $w\!=\!0$), the fixation probability for any
mutant strategy $\mathbf{p}$ simplifies to $\rho(\mathbf{p},\mathbf{q}) =1/N$.
For positive selection strength $w>0$, we thus say that the mutant strategy
$\mathbf{p}$ is advantageous, neutral, or disadvantageous if
$\rho(\mathbf{p},\mathbf{q})$ is larger, equal, or smaller than $1/N$,
respectively. Conversely, we say that the resident strategy $\mathbf{q}$ is
evolutionary robust if there is no advantageous mutant strategy
\cite{stewart:pnas:2013,stewart:pnas:2014}.

For strategy spaces $\mathcal{S}$ with finitely many strategies,
$\mathcal{S}=\{\mathbf{p}^1,\ldots,\mathbf{p}^n\}$, we can use the above formula
for the fixation probabilities to calculate the long-run abundance of each
strategy. For sufficiently rare mutations, the evolutionary process can be
described by a Markov chain with state space $\mathcal{S}$, corresponding
to the homogeneous populations in which everyone applies the same strategy
(see ref.~\citeonline{fudenberg:jet:2006}).
The off-diagonal entries of the transition matrix
$M=(m_{jk})$ are given by $m_{jk}=\rho(\mathbf{p}^k,\mathbf{p}^j)/(n-1)$;
starting in a population in which everyone uses strategy $\mathbf{p}^j$, the
probability that the next mutant adopts strategy $\mathbf{p}^k$ is $1/(n-1)$,
and the probability that the mutant strategy reaches fixation is
$\rho(\mathbf{p}^k,\mathbf{p}^j)$. The diagonal entries of the transition matrix
have the form $m_{jj}=1-\sum_{k\neq j}\rho(\mathbf{p}^k,\mathbf{p}^j)/(n-1)$,
which can be interpreted as the probability that the next mutant strategy will
go extinct. For any finite selection strength $w$, the stochastic transition
matrix $M$ has a unique invariant distribution
$\mathbf{\xi}=(\xi_1,\ldots,\xi_n)$. The entries of $\xi$ represent the
frequency with which each strategy is used in the selection-mutation
equilibrium.
Note that the exact value of the mutation rate is unimportant in
calculating the invariant distribution as long as the transition matrix $M$ is
positive definite.
Using this invariant distribution $\mathbf{\xi}$, one can compute the average
payoff in the population over time as
\begin{equation}
\pi = \sum_j^n \xi_j \cdot \pi(\mathbf{p}^j,\mathbf{p}^j).
\label{e.pay}
\end{equation}
Similarly, one can compute the population's average cooperation rate as
\begin{equation}
\gamma = \sum_j^n \xi_j \cdot \gamma(\mathbf{p}^j,\mathbf{p}^j).
\label{e.rate}
\end{equation}
These two expressions average over all self-interactions of strategies, because
in the rare-mutation limit the population is almost always homogeneous. The
measure $\gamma$ takes into account how much each strategy actually contributes
to the cooperative behaviour of a population. A strategy's contribution may not
always be clear from its definition. For example, the strategy $GT=(1,0,0,0)$ is
a self-defector (as any defection by mistake will cause it to respond with
indefinite defection), whereas $WSLS=(1,0,0,1)$ is a self-cooperator, although
the two strategies differ by just one bit.

When the strategy space is infinite (as for stochastic strategy spaces), we
cannot apply the previous method directly. Instead, we use two different
approximations. The first approach is to discretise the state spaces
$\mathcal{S}=\hat{M}_{1/2}$ and $\mathcal{S}=\hat{M}_1$. That is, instead of
allowing for arbitrary conditional cooperation probabilities $p_{ij}\in[0,1]$,
the probabilities are restricted to some finite grid
$p_{ij}=\{0,1/m,2/m,\ldots,1\}$, where $1/m$ is the grid size. As our second
approach, we use the method of Imhof \& Nowak \cite{imhof:prsb:2010}. This
method starts with an arbitrary resident strategy $\mathbf{p}^{(0)}$. This
resident is then challenged by a single mutant with strategy $\mathbf{q}$, with
$\mathbf{q}$ being taken from a uniform distribution over the space of all
memory-one strategies. If the mutant goes extinct, we define
$\mathbf{p}^{(1)}=\mathbf{p}^{(0)}$; otherwise, the mutant becomes the new
resident and $\mathbf{p}^{(1)}=\mathbf{q}$. This elementary step is repeated for
$t$ iterations, leading to a sequence of successive resident populations
$(\mathbf{p}^{(0)},\mathbf{p}^{(1)},\ldots,\mathbf{p}^{(t)})$. Using this
approach, we can calculate the average payoff of the population as $\pi=\sum_j^t
\pi(\mathbf{p}^{(j)},\mathbf{p}^{(j)})/t$, and the average cooperation rate as
$\gamma=\sum_j^t \gamma(\mathbf{p}^{(j)},\mathbf{p}^{(j)})/t$. As we will see,
the two complementary approaches give similar results -- provided that the grid
size $1/m$ used for the first method is sufficiently small, and that the number
of iterations $t$ used for the second method is sufficiently large.

\subsection*{Analytical methods in the limit of weak selection}

In addition to the above numerical methods, one can use perturbative methods to
compute exact strategy abundance in the limit of weak selection
\cite{antal:jtb:2009,tarnita:jtb:2009}. For a finite strategy space of size $n$,
the assumption of weak selection implies that each strategy $\mathbf{p}^i$ is
approximately played with probability $1/n$, plus a deviation term that is
proportional to
\begin{equation}
L_i = \frac{1}{n} \sum_{j=1}^n \left( \pi(\mathbf{p}^i,\mathbf{p}^i) + \pi(\mathbf{p}^i,\mathbf{p}^j) - \pi(\mathbf{p}^j,\mathbf{p}^i) - \pi(\mathbf{p}^j,\mathbf{p}^j) \right).
\label{eq.lk}
\end{equation}
When $L_i>0$, we say that the strategy $\mathbf{p}^i$ is favoured by selection.
The analogous quantity for infinite strategy spaces (see also
ref.~\citeonline{tarnita:jtb:2009}) is given by
\begin{equation}
L(\mathbf{p}) = \int \left[\pi(\mathbf{p},\mathbf{p}) +
\pi(\mathbf{p},\mathbf{q}) - \pi(\mathbf{q},\mathbf{p}) -
\pi(\mathbf{q},\mathbf{q}) \right] d\mathbf{q}.
\label{eq.l4}
\end{equation}
In this expression $\int d\mathbf{q}$ is the short-hand notation for the
four-dimensional integral $\int_0^1 \int_0^1 \int_0^1 \int_0^1 dq_{CC} dq_{CD}
dq_{DC} dq_{DD}$, which in most cases needs to be computed numerically (see
Appendix). By looking for maxima of $L(\mathbf{p})$, we can determine the
stochastic strategy that is most favoured by selection in the weak-selection
limit.

\section*{Results}

In the following, we first discuss the dynamics in each of the four considered
strategy spaces separately, and then we compare the resulting cooperation levels
and average payoffs.

\subsection*{Strategy dynamics among the deterministic reactive strategies}

The space of deterministic reactive strategies $\mathcal{M}_{1/2}$ consists of
the four strategies $AllD$, $AllC$, $TFT$, and the somewhat paradoxical
Anti-Tit-for-Tat, $ATFT=(0,1,0,1)$, which cooperates if and only if the
co-player was a defector in the previous round. For any set of parameters, we
can use the methods explained in the previous section to calculate the fixation
probability of a mutant with strategy $\mathbf{q}$ in an otherwise homogeneous
population using strategy $\mathbf{p}$.

\begin{figure}[t]
\includegraphics[width=1\textwidth]{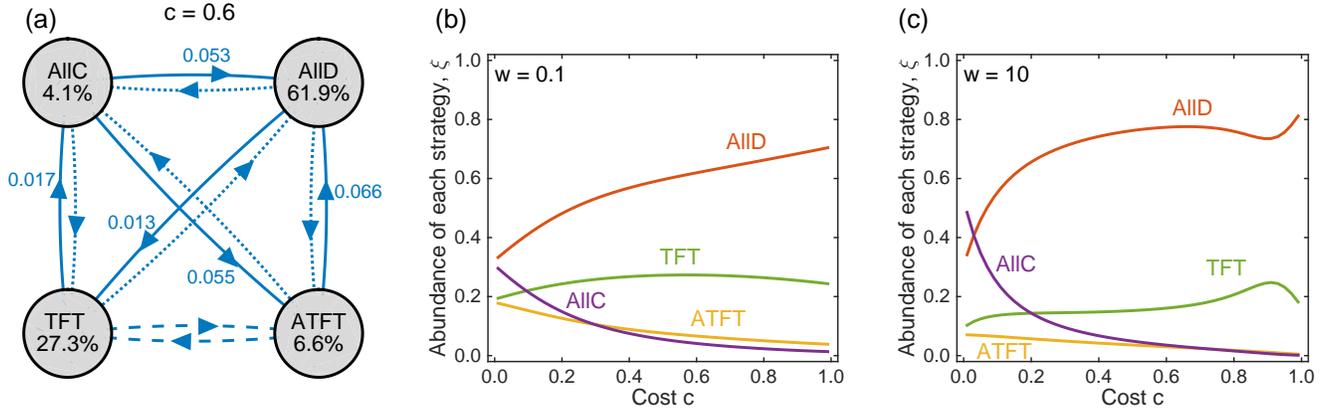}
\caption{Evolutionary dynamics in the space of deterministic reactive
strategies, $\mathcal{M}_{1/2}$. (a) Illustration of the dynamical process. Each
grey circle represents a homogeneous population using one of the four possible
strategies. Blue lines indicate whether a mutant strategy is advantageous (solid
line), neutral (dashed line), or disadvantageous (dotted line). For advantageous
mutants, the blue numbers show the mutant's fixation probability according to
Eq.~(\ref{e.fix}). The graph suggests there are two likely paths for evolution:
a short cycle from $AllD$ to $TFT$ to $AllC$ and back to $AllD$, or the longer
cycle through $AllD$, $TFT$, $AllC$, $ATFT$, and back to $AllD$ (in particular,
eliminating the second cycle by removing $ATFT$ from the strategy set would only
lead to a minor modification of the general dynamics). The numbers within the
grey circles give the abundance of each strategy according to the invariant
distribution of the dynamical process; for the chosen parameters, $AllD$ is the
most abundant strategy. (b) and (c) show the abundance of each strategy
depending on the cost of cooperation and for two different selection strengths
$w=0.1$ and $w=10$. Other parameters: population size $N=100$, error rate
$\varepsilon=0.01$, and in (a) $w=0.1$.}
\label{Fig_PureReact}
\end{figure}

Figure~\ref{Fig_PureReact}a illustrates this procedure in a population of size
$N=100$. If the resident population applies the strategy $AllD$, then neither
$AllC$ nor $ATFT$ are advantageous. A single mutant player with strategy $TFT$,
however, has a fixation probability $\rho=0.013>1/100$ in an $AllD$ population.
$TFT$ can invade because it cannot be exploited
\cite{press:pnas:2012,duersch:ijgt:2013, hilbe:pnas:2014b,hilbe:jtb:2015}: on
average, a $TFT$ player gets the mutual defection payoff $c$ when matched with
an $AllD$-opponent, but it gets $(1+c)/2>c$ when interacting with a
$TFT$-opponent. However, once $TFT$ has reached fixation, a mutant adopting
$AllC$ can easily invade. $AllC$ is more robust to errors -- when two $TFT$
players meet and one player defects by mistake, this can result in long and
costly vendettas between the two players, whereas $AllC$ players would not
encounter that problem. But a homogeneous population of unconditional
cooperators is quickly undermined by defectors, or by $ATFT$ players (who
themselves are typically replaced by defectors). Overall, we end up with an
evolutionary cycle: cooperation can evolve starting from a population of
defectors, but cooperation is not stable.

In the long run, most of the time is spent in a homogeneous $AllD$ population
(for the parameters used in Figure~\ref{Fig_PureReact}a, the abundance of $AllD$
is $61.9\%$). The reason for $AllD$'s predominance is its relative stability: it
takes two $TFT$ players to have a selective advantage in an $AllD$ population (a
single $TFT$ player only obtains the same payoff $c$ that the other $AllD$
players receive). In contrast, it takes only one $AllC$ player to have a
selective advantage in a $TFT$ population, and it takes only one $AllD$ player
to have an advantage in an $AllC$ population. The dynamics within the space of
deterministic reactive strategies is largely independent of the specific
parameters being used. A numerical analysis shows that $AllD$ remains the most
abundant strategy in the selection-mutation equilibrium for both small
(Figure~\ref{Fig_PureReact}b) and large  (Figure~\ref{Fig_PureReact}c) selection
strengths.

We can further confirm these numerical results by analytical means when we look
at the limit of weak selection. For the space $\mathcal{M}_{1/2}$, the linear
coefficients $L_i$ according to Eq.~(\ref{eq.lk}) simplify to
\begin{equation}
\begin{array}{c}
L_{\rm AllD} = c(1-2\varepsilon) > 0\\
L_{\rm TFT} = L_{\rm ATFT} = 0\\
L_{\rm AllC} = -c(1-2\varepsilon) < 0.
\end{array}
\end{equation}
Thus, when selection is weak, $AllD$ is the most abundant strategy for all
values of $c$.

\subsection*{Strategy dynamics among the deterministic memory-one strategies}

\begin{figure}[t]
\includegraphics[width=1\textwidth]{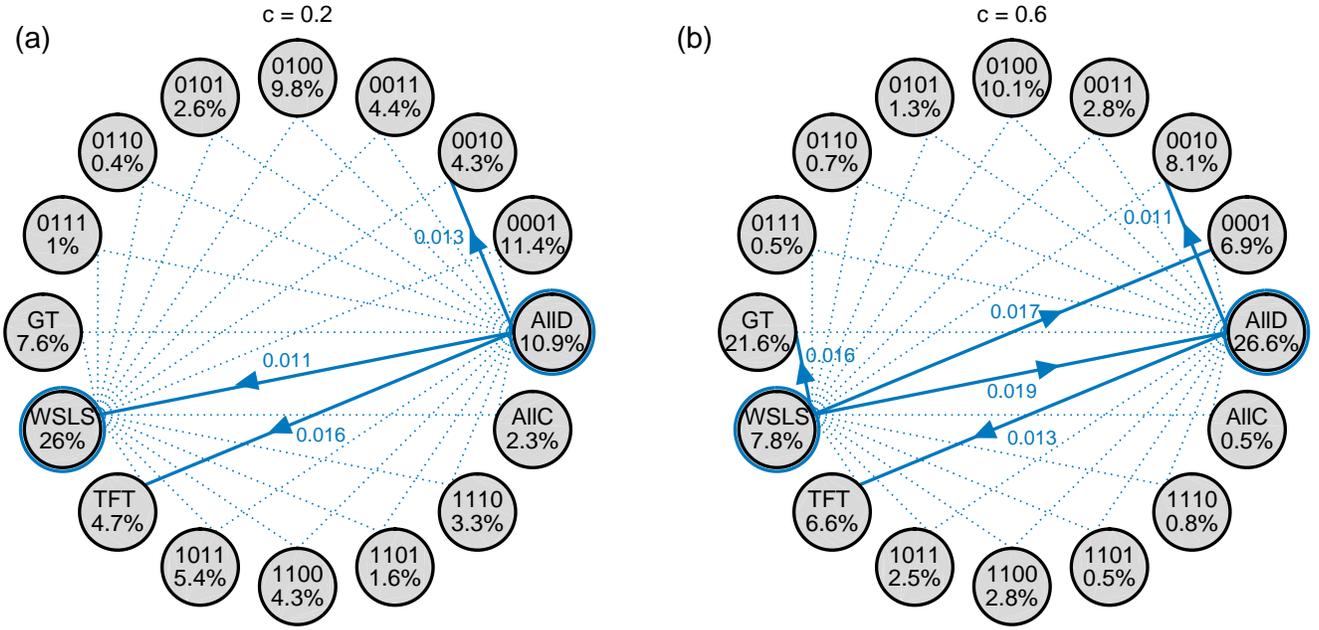}
\caption{Evolutionary dynamics in the space of deterministic memory-one
strategies, $\mathcal{M}_{1}$ for two different cooperation costs. As in
Figure~\ref{Fig_PureReact}a,
the grey circles correspond to all possible homogeneous populations, and
blue lines indicate evolutionary transitions; for clarity, we only show
transitions from $WSLS$ or $AllD$. In (a), the cost of cooperation is
sufficiently low such that $WSLS$ is evolutionary robust. In (b), mutants using
$AllD$, Grim Trigger $GT$, or the strategy $(0,0,0,1)$ can invade a $WSLS$
population; as a consequence, $AllD$ becomes most abundant in the
selection-mutation equilibrium. Parameters are the same as in
Figure~\ref{Fig_PureReact}, population size $N=100$, error rate
$\varepsilon=0.01$, and selection strength $w=0.1$.}
\label{Fig_PureMem1}
\end{figure}

Let us next consider the space of deterministic memory-one strategies, which
contains all 16 tuples of the form ($p_{CC},p_{CD},p_{DC},p_{DD}$) with
$p_{ij}\in \{0,1\}$. Although the state space is now bigger, we can still apply
the previous methods to calculate each strategy's share in the
selection-mutation equilibrium. Figure~\ref{Fig_PureMem1} illustrates two
different parameter scenarios (both assuming an intermediate selection strength,
$w=0.1$). When the costs of cooperation are sufficiently low
(Figure~\ref{Fig_PureMem1}a), the self-cooperating strategy $WSLS$ is
evolutionary
robust: all other mutant strategies have a fixation probability smaller than
$1/N$. In contrast, a population of defectors is not robust: $AllD$ is
susceptible to invasion by $TFT$, $WSLS$, or by the strategy $(0,0,1,0)$. As a
consequence, $WSLS$ is the strategy that is most frequently used over time -- in
the invariant distribution, the share of $WSLS$ is $26.0\%$, whereas the share
of $AllD$ is only $10.9\%$.

\begin{figure}[t!]
\includegraphics[width=0.9\textwidth]{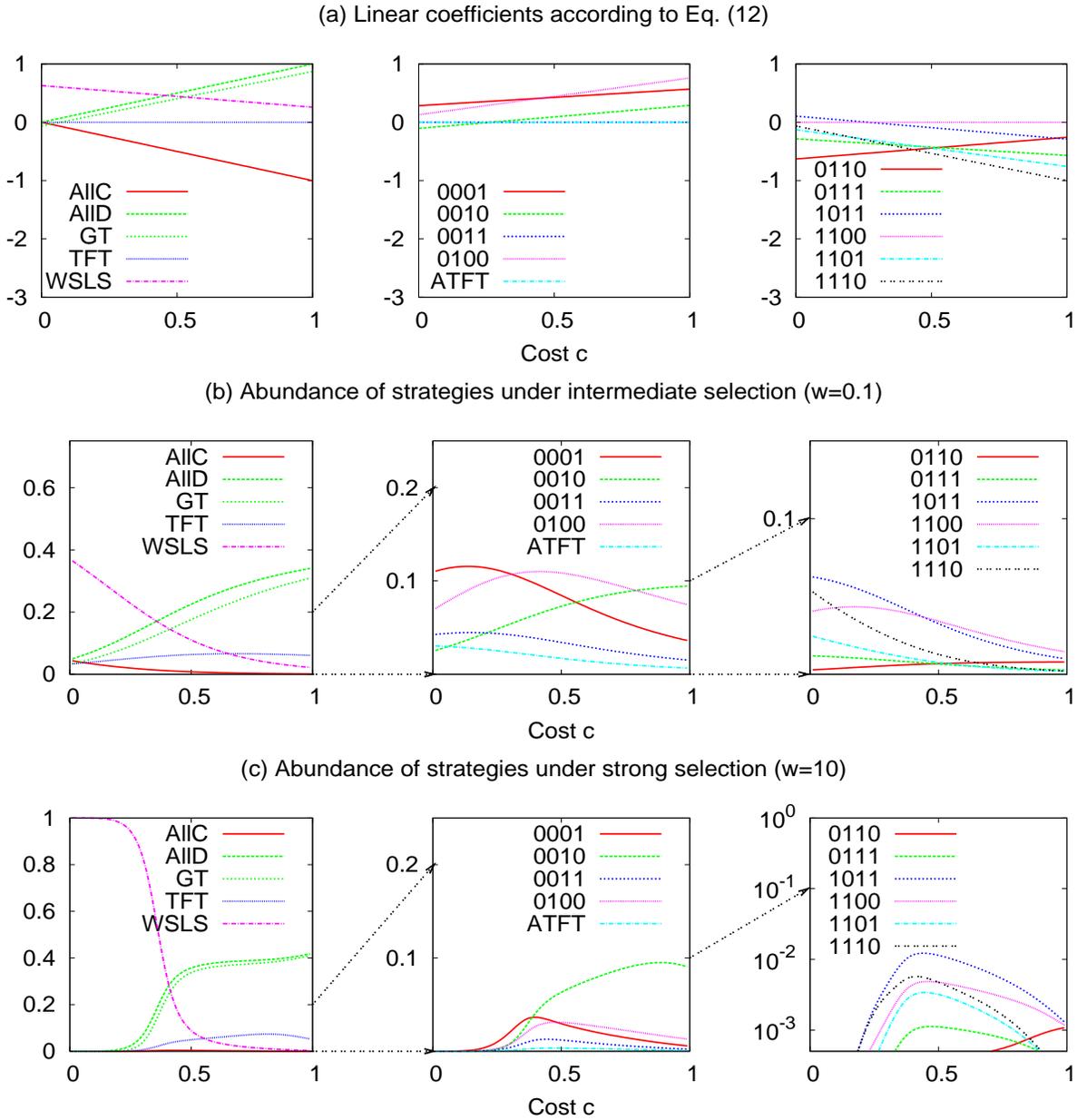}
\caption{Selection-mutation equilibrium in the space of memory-one strategies
for different costs and selection strengths. The graphs in (a) show the linear
coefficients $L_i$ according to Eq.~(\ref{eq.lk}), whereas the graphs in (b) and
(c) show the strategy abundance for intermediate ($w=0.1$) and strong ($w=10$)
selection, respectively. In each case, the 16 curves are plotted in three
different panels (depending on the strategy's abundance), in order to increase
the clarity of the Figure. $WSLS$ is most abundant when cooperation is cheap,
whereas $AllD$ and $GT$ become predominant as $c$ exceeds a critical threshold.
The other parameters are the same as before, $N=100$ and $\varepsilon=0.01$.}
\label{Fig_PureMem1v2}
\end{figure}

The situation changes, however, when the cooperation costs exceed a critical
threshold, as in Figure~\ref{Fig_PureMem1}b. In that case, $WSLS$ ceases to be
evolutionary robust. For example, in a homogeneous population of $WSLS$ players,
playing $WSLS$ yields the mutual cooperation payoff $1$, whereas playing $AllD$
yields the temptation payoff $1+c$ in one round and the mutual defection payoff
$c$ in every other round. Consequently, $AllD$ receives the higher payoff
whenever $c>1/2$. Although $AllD$ is not evolutionary robust either, it now
obtains the largest share in the selection-mutation equilibrium (with $26.6\%$,
as compared to the $7.8\%$ of $WSLS$). Numerical calculations confirm that
$AllD$ becomes the most abundant strategy as the cost-to-benefit ratio
approaches 1/2 (see Figure~\ref{Fig_PureMem1v2}). On the positive side, when
cooperation is relatively cheap and when selection is strong, $WSLS$ can reach
almost 100\% in the selection-mutation equilibrium
(Figure~\ref{Fig_PureMem1v2}c).

Again, we can derive analytical results in the limit of weak selection by
calculating the linear coefficients $L_i$ according to  Eq.~(\ref{eq.lk}). There
are only a handful of strategies for which $L_i>0$ independent of the value of
$c$ (see also Figure~\ref{Fig_PureMem1v2}a). Among these are $AllD$ and $WSLS$,
\begin{equation}
\begin{array}{l}
L_{\rm WSLS} = (151-89c)/240 + \mathcal{O}(\varepsilon),\\
L_{\rm AllD} = c + \mathcal{O}(\varepsilon).
\end{array}
\end{equation}
In particular, $WSLS$ is most abundant when $L_{\rm WSLS}>L_{\rm AllD}$, or equivalently, when $c<c_0:=  151/329 + \mathcal{O}(\varepsilon)\approx~0.46$.

\subsection*{Strategy dynamics among the stochastic reactive strategies}

\begin{figure}[t]
\includegraphics[width=1\textwidth]{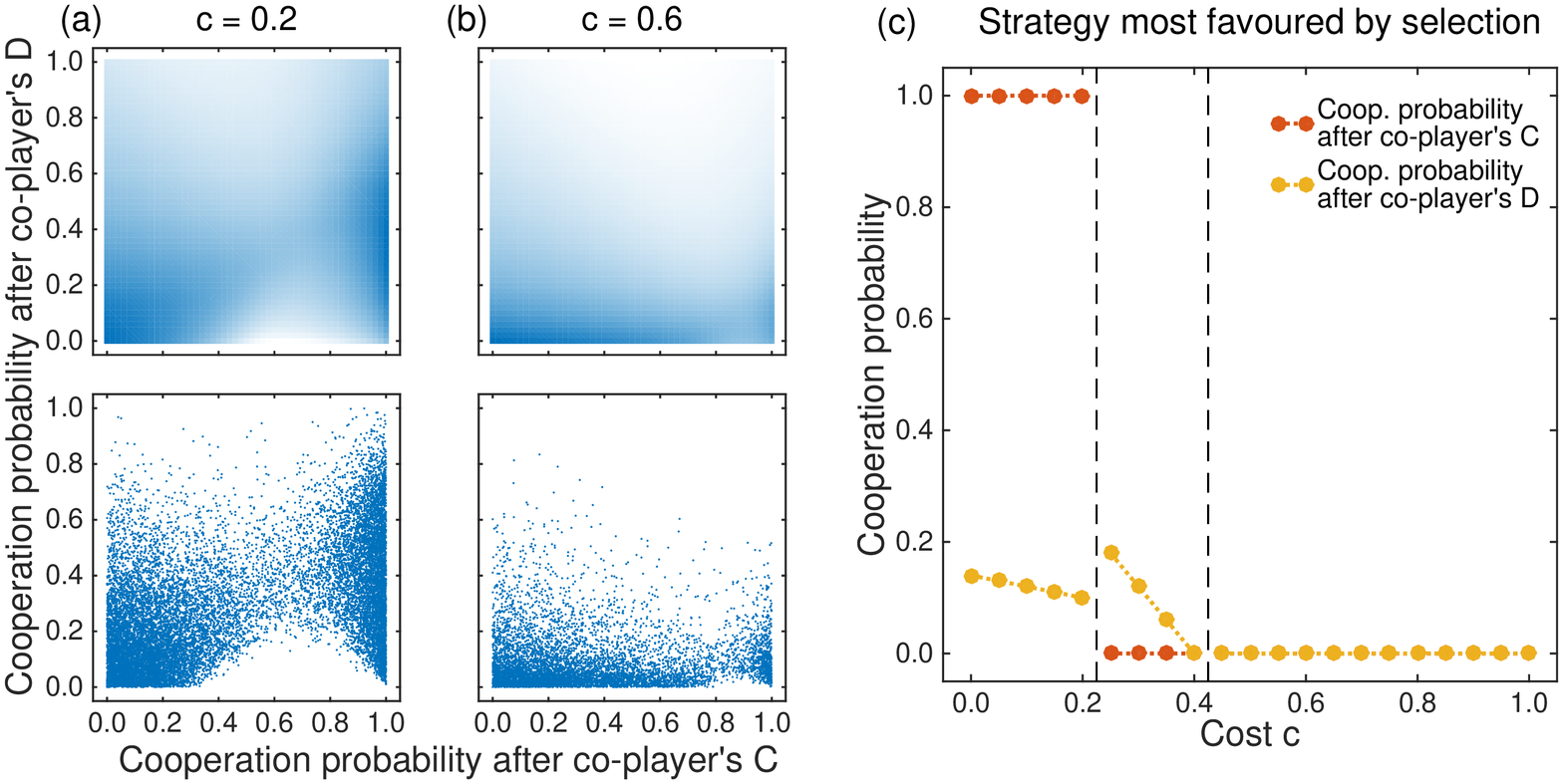}
\caption{Evolutionary dynamics in the space of stochastic reactive strategies,
$\hat{\mathcal{M}}_{1/2}$. (a) and (b) illustrate our approximation for the
invariant distribution for two different cost values, $c=0.2$ and $c=0.6$. For
the upper graphs, we have calculated the invariant distribution for the
discretised state space, where the conditional cooperation probabilities of the
reactive strategy are taken from the (finite) set
$\{0,\delta,2\delta,\ldots,1-\delta,1\}$, using a grid size $\delta=0.02$. Areas
in dark blue colour correspond to strategy regions that have a relatively high
frequency in the invariant distribution. The lower graphs show the results of
simulations for the Imhof-Nowak process \cite{imhof:prsb:2010}; each blue dot
represents a strategy adopted by the resident population. Both methods confirm
that when the cost of cooperation is small, e.g. $c=0.2$, the resident
strategies are either clustered around the lower left corner or around the right
edge of the state space. As the cost increases, more weight is given to the
lower edge. In (c) we show the strategy that is most favoured in the limit of
weak selection, i.e., the strategy with the highest linear coefficient
$L(\mathbf{p})$ according to Eq.~(\ref{eq.l4}). The graph indicates that there
are three parameter regions: for low cost values, a generous strategy is most
favoured; for intermediate cost values, the most favoured strategy has only a
positive cooperation probability if the co-player defected previously; and for
high cooperation costs $AllD$ is most favoured. Parameters: Population size
$N=100$, $\varepsilon=0.01$, and $w=10$; the Imhof-Nowak process was simulated
over $5\cdot 10^6$ mutant strategies.}
\label{Fig_StochReact}
\end{figure}

Let us next turn to stochastic reactive strategies. In that case, players only
pay attention to the co-player's previous move (i.e., $p_{CC}=p_{DC}$ and
$p_{CD}=p_{DD}$), but now they are able to choose their cooperation
probabilities from the unit interval, $p_{ij}\in [0,1]$. In particular, there
are now infinitely many feasible strategies, which renders a full calculation of
all transitions between possible homogeneous populations impossible. To cope
with this issue, we have used two numerical approximations. The first method
approximates the infinite state space by a finite grid (to which the previously
used methods for finite strategy spaces can be applied). For two different cost
values, we have illustrated the resulting invariant distribution in the upper
panels of Figure~\ref{Fig_StochReact}a and  \ref{Fig_StochReact}b.
Figure~\ref{Fig_StochReact}a indicates that when cooperation costs are low,
there are
two strategy regions with a high abundance according to the invariant
distribution. The first region corresponds to a neighbourhood of $AllD$ (i.e.
strategies for which both conditional cooperation probabilities are low); the
second region comprises a set of generous strategies. In that region, players
always reciprocate their opponent's cooperation, while still exhibiting some
degree of forgiveness in case the opponent defected in the previous round.
However, as the cooperation costs increase (as in Figure~\ref{Fig_StochReact}b
for which $c=0.6$), the region of generous strategies is visited less often, and
defective strategies become predominant.

We obtain a similar result when we use our second method to approximate the
dynamics within the space of stochastic reactive strategies. For this method, we
have applied the dynamics of Imhof \& Nowak \cite{imhof:prsb:2010}: starting
from a population of defectors, we have repeatedly introduced single mutants
into the population, who may adopt an arbitrary stochastic strategy (i.e., this
time, strategies are not restricted to some finite grid). The mutant strategy
may then either fixate or go extinct, leading to a sequence of resident
populations over time. The lower panels in \ref{Fig_StochReact}a and
\ref{Fig_StochReact}b depict the residents in this sequence as blue dots (for
clarity, we have only plotted those resident populations that survived at least
50 mutant invasions). Again, low cooperation costs lead to two clusters in the
two-dimensional state space -- a cluster with defective strategies and a cluster
with generous strategies. But as before, the cluster of generous strategies
tends to shrink as the cooperation costs increase (as also observed in
ref.~\citeonline{imhof:prsb:2010}).
We have also numerically computed the stochastic reactive strategy that is most
favoured by selection (see Figure~\ref{Fig_StochReact}c). There are three
parameter regions: for cost-to-benefit ratios below 1/4, we observe that the
most favoured strategy is generous. However, as the cooperation costs increase
and the cost-to-benefit ratio is between 1/4 and 2/5, the most favoured strategy
prescribes that players should no longer reciprocate cooperation, and players
should only cooperate with some low probability when the opponent defected in
the previous round. Clearly, a population made up of such players only achieves
low levels of cooperation. The situation becomes even worse as the
cost-to-benefit ratio exceeds 2/5, in which case unconditional defection becomes
the most favoured strategy.

\subsection*{Strategy dynamics among the stochastic memory-one strategies}

\begin{figure}[t]
\includegraphics[width=0.8\textwidth]{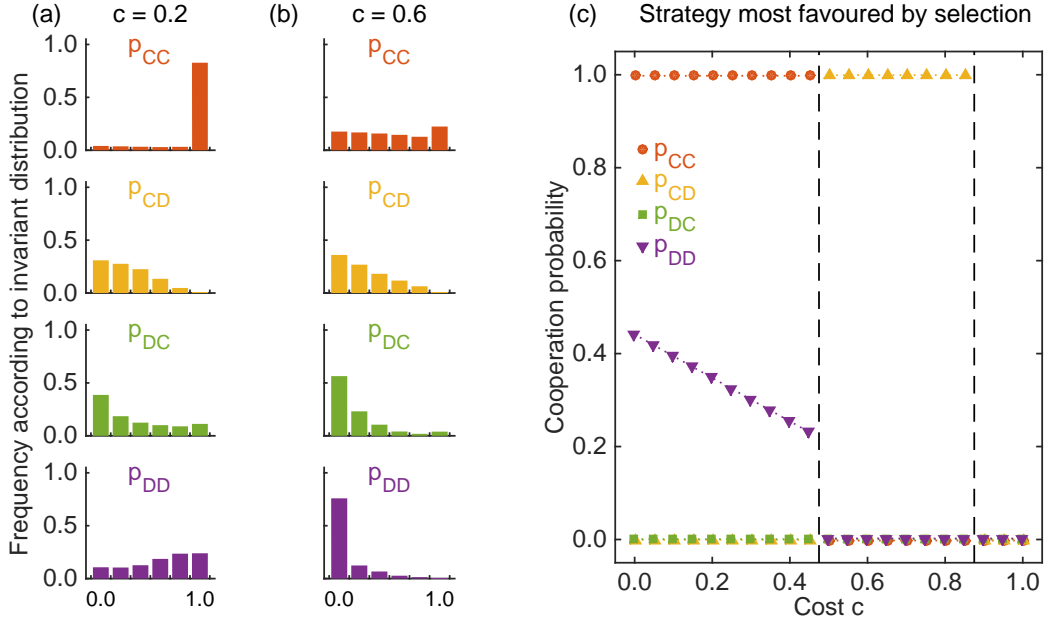}
\caption{Evolutionary dynamics in the space of stochastic memory-one strategies,
$\hat{\mathcal{M}}_{1}$. (a) and (b) show the marginal distribution of the
evolving cooperation probabilities $p_{ij}$ in the mutation-selection
equilibrium. To generate the figure, we have calculated the invariant
distribution for a discretised version of the state space, using a grid size of
$\delta=0.2$. For low costs, the cooperation probabilities are in line with
$WSLS$ behaviour; for larger cost values, cooperation breaks down, and most
evolving strategies are self-defectors. In (c) we depict the strategy that has
the highest linear coefficient $L(\mathbf{p})$ according to Eq.~(\ref{eq.l4}).
Again there are three parameter regions: for low costs, a variant of $WSLS$ is
most favoured by selection; for intermediate costs, the somewhat paradoxical
strategy $(0,1,0,0)$ is most favoured; and for high costs, $AllD$ becomes
predominant. Parameters are the same as before: Population size $N=100$,
$\varepsilon=0.01$, and $w=10$.}
\label{Fig_StochMem1}
\end{figure}

Finally, we can apply the same two approximations to the 4-dimensional space of
all stochastic memory-one strategies. Of course, that state space can no longer
be depicted in a two-dimensional graph; but Figures~\ref{Fig_StochMem1}a and
\ref{Fig_StochMem1}b show the invariant distribution for each of the four
components $p_{CC}$, $p_{CD}$, $p_{DC}$ and $p_{DD}$, again for the two cost
values $c=0.2$ and $c=0.6$. For $c=0.2$ we observe behaviour that is consistent
with $WSLS$. After mutual cooperation, players almost certainly continue with
cooperation, and after mutual defection players are more likely to cooperate
than to defect, whereas the values of $p_{CD}$ and $p_{DC}$ rather prescribe to
defect in the next round. On the other hand, when $c=0.6$, the invariant
distribution shows a bias towards self-defector strategies, as mutual defection
in one round is most likely to lead to mutual defection in the next round.
Again, we have also calculated the strategy most favoured by selection in the
limit of weak selection (Figure~\ref{Fig_StochMem1}c). As in the case of
stochastic reactive strategies, there are three scenarios: a cooperative
scenario in which the population applies a variant of $WSLS$ when cooperation
costs are low; an intermediately cooperative scenario where the population uses
the strategy $\mathbf{p}^*=(0,1,0,0)$; and a defection scenario of an $AllD$
population when cooperation costs are high. Compared to the case of reactive
strategies, the fully cooperative strategy is now favoured for a wider range of
cost values -- the $WSLS$ variant is most abundant for costs $c\lesssim 0.45$,
whereas the $GTFT$-like strategy depicted in Figure~\ref{Fig_StochReact}c can
only succeed when $c\lesssim 0.25$. $WSLS$ variants of the form $(1,0,0,x)$ have
the advantage of being immune against the invasion by both, $AllC$ and $AllD$
mutants (provided that $x$ is sufficiently small for given cooperation costs).
However, as opposed to the pure WSLS strategy $(1,0,0,1)$, strategies of the
form $(1,0,0,x)$ with $x<1$ are not evolutionary robust. In the presence of
errors, they can be invaded by strategies that yield a better approximation to
$WSLS$, (1,0,0,y) with $y>x$, which in turn are more susceptible to invasion by
$AllD$. As a consequence, we observe that the parameter region for which $WSLS$
variants are most favoured in the space of stochastic memory-one strategies is
comparable to the region for which the pure $WSLS$ strategy is most abundant
among the deterministic strategies (as depicted in Figure~\ref{Fig_PureMem1v2}).

Among the strategies most favoured by selection, the strategy
$\mathbf{p}^*=(0,1,0,0)$ comes most unexpected~\cite{kim:jkps:2014}.
This strategy prescribes to
cooperate only if one has been exploited in the previous round -- which seems to
be a rather paradoxical response. For small errors, a homogeneous population of
$\mathbf{p}^*$ players yields an expected payoff of $\pi^*=(1+3c)/4$; two
$\mathbf{p}^*$-players would typically defect against each other, but if one of
the player cooperates by error, there can be long periods of unilateral
cooperation. However, a single mutant applying $AllD$ obtains the higher payoff
$(1+3c)/3$, and thus one would expect that homogeneous $\mathbf{p}^*$
populations quickly disappear. But if $\mathbf{p}^*$ is not evolutionary robust,
how can it be most favoured by selection for intermediate cost ranges?

Although $AllD$ could easily invade a $\mathbf{p}^*$--population, it is highly
unlikely that within the space of stochastic memory-one strategies the next
mutant actually adopts $AllD$. Instead, most arising mutants would use
strategies $\mathbf{p}=(p_{CC},p_{CD},p_{DC},p_{DD})$ for which all cooperation
probabilities $p_{ij}$ are strictly positive. In the limit of small errors,
$\varepsilon \rightarrow 0$, the payoff of such mutants in a
$\mathbf{p}^*$--population can be computed as
\begin{equation}
\pi=\frac{1-p_{CD}}{1-p_{CD}+p_{DD}}c.
\end{equation}
This payoff is not only smaller than the residents' payoff $\pi^*$; it is
exactly the same payoff that mutants would get in an $AllD$ population. Thus,
the strategy $\mathbf{p}^*=(0,1,0,0)$ can be successful because against almost
all mutant strategies it behaves like $AllD$; only against itself (and against a
few other strategies, like against $AllD$) it cooperates occasionally. In a
sense, $\mathbf{p}^*$ acts as if it used a rudimentary form of kin recognition -
it shows some cooperation against players of the same kind, but it defects
against almost everyone else.

\subsection*{Comparison of the evolving cooperation rates}

\begin{figure}[t]
\includegraphics[width=\textwidth]{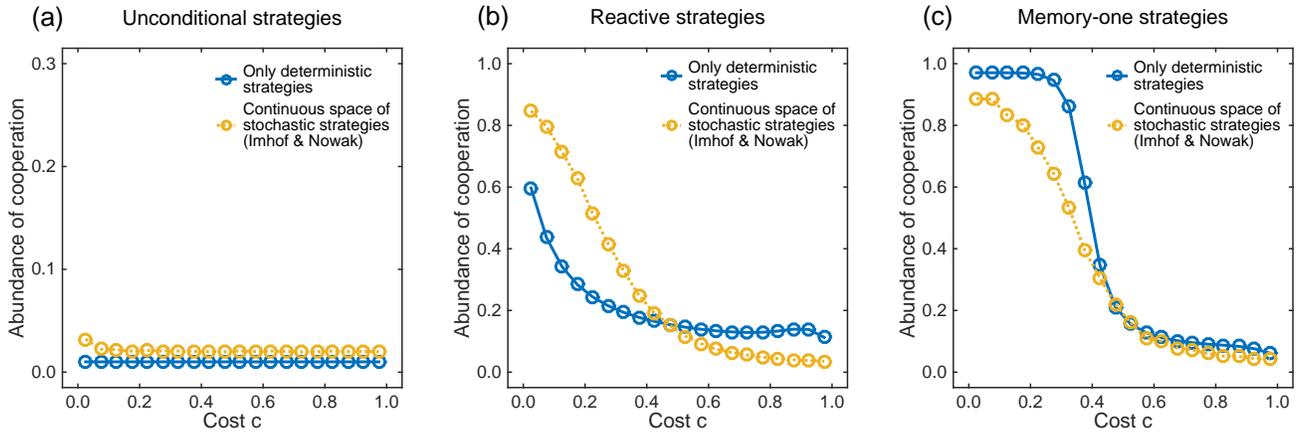}
\caption{Evolving cooperation rates for (a) unconditional strategies, (i.e.,
strategies that use the same cooperation probability $p$ in every round,
independent of the past history), (b) reactive strategies, and (c) memory-one
strategies. All graphs show the abundance of cooperation as measured by the
quantity $\gamma$ in Eq.~(\ref{e.rate}) for the case of deterministic strategies
(blue), and according to the Imhof-Nowak process for stochastic strategies
(yellow; a discretised version of the continuous space of memory-one strategies
would yield similar results). Dots represent simulation results, whereas solid
lines represent numerically exact results derived from the invariant
distribution of the evolutionary processes. Parameters: population size $N=100$,
$\varepsilon=0.01$, and $w=10$.}
\label{Fig_CoopRate}
\end{figure}

After analysing the strategy dynamics in each of the four strategy spaces
separately, we are now in a position to compare the evolving cooperation rates.
For reactive strategies and low cooperation costs, stochastic strategies lead to
more cooperation than deterministic strategies (Figure~\ref{Fig_CoopRate}b). As
we have seen in Figure~\ref{Fig_PureReact}, deterministic reactive strategies
are unable to stabilize cooperation; $TFT$ can be invaded by $AllC$, and $AllC$
is easily invaded by $AllD$ (see also ref.~\citeonline{imhof:pnas:2005}).
Stochastic
reactive strategies, on the other hand, can maintain a healthy level of
cooperation for a considerable time. $GTFT$-like strategies resist invasion by
$AllD$, and they are only destabilized when altruistic $AllC$-like strategies
increase in frequency by neutral drift
\cite{nowak:aam:1990,nowak:nature:1992,imhof:prsb:2010,grujic:jtb:2012,hilbe:plosone:2013b,dong:plosone:2015}.
However, with increasing cooperation costs, it takes longer until $GTFT$-like
strategies emerge, as the so-called cooperation-rewarding zone shrinks as $c$
increases
(see, for example, ref.~\citeonline{sigmund:book:2010}), and $GTFT$-like
strategies are more likely to be invaded by overly altruistic strategies.  As a
result, when cooperation costs are high deterministic strategies perform
slightly better, because $TFT$ mutants show up more quickly to re-invade $AllD$
populations.

Memory-one strategies are generally more favourable to cooperation, as depicted
in Figure~\ref{Fig_CoopRate}c. In contrast to reactive strategies, memory-one
strategies allow for $WSLS$-like behaviour which is more stable against indirect
invasion by altruistic $AllC$ strategies
\cite{nowak:nature:1993,imhof:jtb:2007}. Interestingly, however, we find that
for low cooperation costs, deterministic memory-one strategies are better in
sustaining cooperation than stochastic strategies. Among the deterministic
memory-one strategies, mutants are strongly opposed by selection when they enter
a $WSLS$ population (as illustrated in Figure~\ref{Fig_PureMem1}). As a result,
$WSLS$ reaches almost $100\%$ in the invariant distribution, provided that
selection is sufficiently strong and that the costs of cooperation are low.
There are two reasons why stochastic strategies can result in less cooperation.
First, although $WSLS$ remains a Nash equilibrium
\cite{akin:2013,hilbe:geb:2015}, stochasticity allows for the invasion of nearby
mutants (that are only slightly disfavoured by selection); these mutants may in
turn be more susceptible to invasion by $AllD$ \cite{garcia:jet:2016}. Second,
stochastic dynamics often generates resident populations that only use an
approximate version of $WSLS$, having the form $(1,0,0,x)$, with $x<1$. Compared
to the deterministic $WSLS$ rule, these approximate versions are more prone to
noise: if one of the players defected by error, it may take a substantial number
of rounds to re-establish mutual cooperation (which becomes most clear when $x$
is close to zero).

\begin{figure*}[t]
\begin{center}
\includegraphics[width=0.8\textwidth]{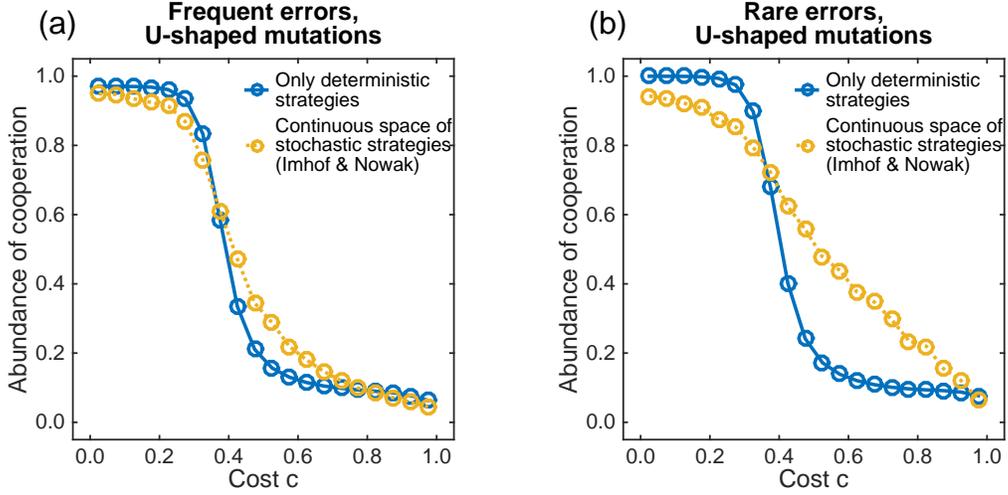}
\caption{U-shaped mutation kernels lead to more cooperation in high cost
scenarios. As in Figure~\ref{Fig_CoopRate}c, both graphs show the evolving
cooperation rate for the space of deterministic memory-one strategies (blue) and
stochastic memory-one strategies (yellow). However, here we have
varied the error rate of players ($\varepsilon=1\%$ for frequent errors,
$\varepsilon=0.01\%$ for rare errors). In addition, the cooperation
probabilities $p_i$ of new mutant strategies are now taken from a
beta-distribution. The beta-distribution has the density function
$f(p)=Cp^{\alpha-1}(1-p)^{\beta-1}$, with $C$ being a normalization factor. The
values $\alpha=\beta=1$ yield the uniform distribution on [0,1], as used in
Figure~\ref{Fig_CoopRate}; here, we have taken $\alpha=\beta=0.1$, yielding a
strongly U-shaped distribution. All other parameters are the same as in
Figure~\ref{Fig_CoopRate}.} \label{Fig_UShape}
\end{center}
\end{figure*}

This result is somewhat disappointing: especially in parameter regions in which
$WSLS$ is unstable, one would hope that stochastic strategies allow at least for
some degree of cooperation, because WSLS variants of the form $(1,0,0,x)$ are
immune to the invasion of $AllC$ and $AllD$ mutants as explained above.
The previous results on the effect of stochasticity
need to be viewed in light of the assumed mutation kernel -- for our numerical
results we have assumed that new mutant strategies are taken from a uniform
distribution. This assumption often generates mutant strategies with
intermediate cooperation probabilities -- which have no chance of being
evolutionary robust \cite{stewart:pnas:2014}. What would happen if mutant
strategies were instead taken from a distribution that puts more weight on the
boundary of the state space? In Figure~\ref{Fig_UShape}, we show numerical
results under the assumption that the cooperation probabilities of new mutant
strategies follow a U-shaped distribution on the interval [0,1]. Keeping the
previous error rate of $\varepsilon=0.01$, the U-shaped mutation kernel seems to
marginally increase the evolving cooperation rates for most cost values
(Figure~\ref{Fig_UShape}a). If we additionally reduce the error rate to
$\varepsilon=10^{-4}$, U-shaped mutations can lead to a more dramatic increase
in cooperation rates, especially for scenarios with intermediate cooperation
costs. In that parameter region, successful residents often apply strategies of
the form $(1-\delta_1,\delta_2,\delta_3,\delta_4)$ with all $\delta_i \ll 1$.
Because $\delta_4\ll 1$, such residents can hardly be exploited by $AllD$
mutants. If, in addition, $\delta_1 \ll \delta_4$, such strategies can still
reach a substantial level of cooperation against themselves. We note that
strategies of the form $(1-\delta_1,\delta_2,\delta_3,\delta_4)$ are not stable,
as they could be invaded by strategies that increase their cooperation
probability after mutual defection. However, provided that $\delta_1$ is
sufficiently small, the selective advantage of such mutants would be comparably
small, and hence it may take a long time until such mutant strategies appear and
fixate in the population. The results in Figure~\ref{Fig_UShape} thus suggest
that the assumed mutation structure can have a considerable impact on the
evolving cooperation rates. Herein, we have considered two extreme structures,
uniform mutations and strongly U-shaped mutations, but a more general analysis
of the impact of different mutation kernels would certainly be a worthwhile
topic for future research.

\section*{Discussion and Summary}

We have used the Moran process in finite populations to study the evolution of
cooperation in repeated games. The mathematics of repeated games can be
intricate. Even if one only considers a restricted strategy space, such as the
space of all memory-one strategies, it is typically hard to derive exact results
for the resulting evolutionary dynamics. There are various ways to cope with
this complexity. Some studies have focused on even simpler strategy sets,
consisting only of a handful of representative strategies (e.g.
refs.~\citeonline{boyd:jtb:1988,imhof:pnas:2005,grujic:jtb:2012,van-segbroeck:prl:2012}).
Others have obtained analytical results for certain infinitely-dimensional
subsets of memory-one strategies, like reactive strategies \cite{nowak:aam:1990,
imhof:prsb:2010}, zero-determinant strategies \cite{hilbe:plosone:2013b}, or
conformistic strategies \cite{dong:plosone:2015}. Yet another approach is to use
computer simulations (as in refs.~\citeonline{nowak:nature:1993,
hilbe:pnas:2013, solnoki:pre:2014, stewart:games:2015}).
Herein, we have taken a somewhat
intermediate approach. By assuming appropriate separation of time scales
(e.g., mutations are sufficiently rare such that populations are typically
homogeneous), we can compute numerically exact strategy abundance in case the
strategy space is finite (as in the case of deterministic strategies). To
explore the dynamics among stochastic strategies, we have extended this approach
to approximate the dynamics in infinite strategy spaces.

We have used this approach to systematically compare the evolutionary dynamics
among strategy spaces of different complexity. The strategy spaces considered
differ along two dimensions, depending on whether strategies are reactive or
memory-one, and depending on whether strategies are deterministic or stochastic.
Each of the four considered strategy spaces has been explored previously, but
only in isolation. Herein, we are explicitly interested how much complexity is
needed to allow for a healthy level of cooperation. In this way, our study
contributes to a growing research effort, exploring how the evolution of
cooperation depends on underlying modelling assumptions. For example,
Garc{\'i}a and Traulsen \cite{garcia:plosone:2012} and Stewart and Plotkin
\cite{stewart:games:2015} have analysed the role of the mutation structure on
the emergence and stability of cooperation, whereas van den Berg and Weissing
\cite{vandenberg:prsb:2015} have explored the consequences of two different
strategy representations. We believe that this kind of research is extremely
useful, as it serves as an important robustness check for previous results on
the evolution of direct reciprocity.

Our study provides at least two major insights. The first insight is that more
complex strategies do not guarantee more cooperation. More specifically, we have
found that memory-one strategies, which also take one's own previous move into
account, have a positive impact on
cooperation. If players have no memory at all (i.e. if they can only use
unconditional strategies), evolution unambiguously promotes defection (as
depicted in Figure~\ref{Fig_CoopRate}a). However, if players can react to the
co-player's previous move, or even better to the moves of both players, then
evolution can promote cooperative strategies when the costs of cooperation are
sufficiently low. Although we have not tested memory-two strategies (i.e.
players who react to the outcome of the last two rounds), one may expect that
such strategies could further facilitate cooperation, especially in parameter
regions in which the classical $WSLS$ strategy becomes unstable (see, e.g.
refs.~\citeonline{hauert:prsb:1997,baek:pre:2008}). The effect of stochasticity
on cooperation is more ambiguous. If players only remember the co-players'
previous move, then stochasticity allows for generous strategies like $GTFT$,
and such generous strategies can help to establish relatively high levels of
cooperation.  On the other hand, when cooperation costs are low, and players are
allowed to use memory-one strategies, stochastic strategies cannot further
promote cooperation. Here, the deterministic version of $WSLS$ works best.

Our second insight is rather conceptual. To quantify the evolutionary success of
some strategy $\mathbf{p}$, it is common to check whether the strategy is an
equilibrium, or whether the strategy is evolutionary robust (see e.g.
refs.~\citeonline{boyd:nature:1987,stewart:pnas:2013,akin:2013,stewart:pnas:2014,hilbe:geb:2015}).
To this end, one checks whether there would be a mutant strategy $\mathbf{q}$
that can prosper in a population of $\mathbf{p}$ players. A strategy that is not
robust is generally assumed to play a minor role during the evolutionary
process. Yet, we have seen that under some evolutionary conditions, the strategy
$\mathbf{p}^*=(0,1,0,0)$ can be surprisingly successful despite not being
evolutionary robust. This somewhat paradoxical strategy can persist because
against almost all other strategies it plays like $AllD$; but against a handful
of strategies (including itself and against $AllD$) it cooperates for a
substantial fraction of time. In particular, there are mutant strategies that
could invade into a homogeneous $\mathbf{p}^*$~-~population. However, the
probability that such a mutant arises within a reasonable timespan is
vanishingly small, as the space of such advantageous mutants has measure zero
within the space of all memory-one strategies. Thus, it does not seem sufficient
for evolutionary robustness to ask whether there is another strategy that would
have a higher fitness; one also needs to check whether this beneficial mutant
strategy can arise under the considered mutation scheme. Put differently, unless
a resident outperforms every other strategy, the question of evolutionary
robustness cannot be properly assessed without reference to the mutation scheme.
Of course, this observation does not diminish the value of
traditional equilibrium considerations -- but if a strategy is only unstable
because some non-generic strategy can invade, then some caution seems warranted.

\section*{Appendix: Computation of the linear coefficient $L(\mathbf{p})$
for stochastic strategies}

To compute the stochastic strategy that is most favoured by selection, we have
evaluated the four-dimensional integral $L(\mathbf{p})$ in Eq.~(\ref{eq.l4}) by
means of Gaussian quadrature~\cite{newman:book:2013}. For maximizing
$L(\mathbf{p})$, we have employed a two-step approach: The first step is
exhaustive global search of the whole strategy space. Some degree of
discretisation is inevitable in checking many different realizations of
$\mathbf{p} = (p_{CC}, p_{CD},p_{DC}, p_{DD})$. In particular, we have observed
that the objective function $L(\mathbf{p})$ tends to change rapidly when
$\mathbf{i}$ approaches the boundary of the strategy space. As the change is
smoothed by the implementation error, it is quite often the case that $p'_{ij}$,
or $1-p'_{ij}$, turns out to be $\mathcal{O}(\varepsilon)$. Therefore, the mesh
size of $p_{ij}$ has been set to be of an order of $\varepsilon$ when getting
close to zero or one. Specifically, we have used $17^4 = 83,521$ grid points in
total by adding $p_{ij} = 0.005, 0.01, 0.02, 0.98, 0.99$, and $0.995$ to a
regular mesh grid $p_{ij} = 0.1 k$ ($k=0, \ldots, 10$).

The next step is the gradient-descent method~\cite{press:book:1992}, starting
from the best strategy of the exhaustive search. Although this second method is
local, it works in a continuous space and finds out a nearby maximum with far
higher precision than the grid search. We expect that this two-step approach
precisely locates the global maximum as long as the mesh of the first step is
fine enough to detect all the relevant variations of the objective function
$L(\mathbf{p})$.


\section*{Acknowledgements}
S.K.B. gratefully acknowledge discussions with Su Do Yi. S.K.B. was supported by
Basic Science Research Program through the National Research Foundation of Korea
(NRF) funded by the Ministry of Science, ICT and Future Planning
(NRF-2014R1A1A1003304). H.C.J. was supported by Basic Science Research Program
through the National Research Foundation of Korea (NRF) funded by the Ministry
of Education (2015R1D1A1A01058317). C.H. acknowledges generous funding from the
Schr\"{o}dinger scholarship of the Austrian Science Fund (FWF), J3475.

\section*{Author contributions statement}

H.-C.J. and M.A.N. designed the research, S.K.B. and C.H. performed the
simulations and analysed the results. All authors wrote and reviewed the
manuscript.

\section*{Additional information}

Competing financial interests: The authors declare no competing financial
interests.

\end{document}